\documentclass[12pt]{article}
\usepackage{graphicx, epstopdf, amsmath, amssymb, amsfonts, comment}

\newcommand{\R}{\mathbb{R}} 
\newcommand{\E}{\mathbb{E}}

\newtheorem{definition}{Definition}[section]
\newtheorem{theorem}{Theorem}[section]

\newtheorem{proposition}{Proposition}[section]
\newtheorem{lemma}{Lemma}[section]
\newtheorem{remark}{Remark}[section]

\begin{document}

\title{Analysis of the nonlinear option pricing model under variable transaction costs}

 
\author{Daniel \v{S}ev\v{c}ovi\v{c}
\thanks{
Department of Applied Mathematics and Statistics, Comenius University, 842 48 Bratislava, Slovakia, sevcovic@fmph.uniba.sk
}
\and Magdal\'ena \v{Z}it\v{n}ansk\'a
\thanks{
Department of Mathematics and Actuarial Science, University of Economics in Bratislava,  852 35 Bratislava, Slovakia, magdalena.zitnanska@euba.sk
\newline
The research was supported by FP7-PEOPLE-2012-ITN project \# 304617 STRIKE and APVV--SK-PT-0009-12  grants.
}
}


\date{}
\maketitle

\begin{abstract}

In this paper we analyze a nonlinear Black--Scholes model for option pricing under variable transaction costs. The diffusion coefficient of the nonlinear parabolic equation for the price $V$ is assumed to be a function of the underlying asset price and the Gamma of the option. We show that the generalizations of the classical Black--Scholes model can be analyzed  by means of transformation of the fully nonlinear parabolic equation into a quasilinear parabolic equation for the second derivative of the option price. We show existence of a classical smooth solution and prove useful bounds on the option prices. Furthermore, we construct an effective numerical scheme for approximation of the solution. The solutions are obtained by means of the efficient numerical discretization scheme of the Gamma equation. Several computational examples are presented. 
\end{abstract}



\section{Introduction}

The classical linear Black--Scholes option pricing model with a constant historical volatility was proposed in \cite{BS}. The model was derived  under several restrictive assumptions, such as assumption on market completeness, continuous trading and zero transaction costs. According to this option pricing theory the price $V(S,t)$ of a contingent claim written on the underlying asset $S>0$ at the time $t\in[0,T]$  is a solution to the linear parabolic equation 
\begin{equation} \label{linBS}
\partial_ t V+\frac{1}{2}\sigma^2S^2\partial^2_
S V+r S \partial_S
V-rV=0,
\end{equation}
where $r > 0 $ is the risk-free interest rate of a zero-coupon bond and $\sigma$ is the historical volatility of the underlying asset which is assumed to follow a stochastic differential equation of the geometric Brownian motion, i.e. 
\begin{equation}
d S = \rho S \, dt + \sigma S \, d W,
\end{equation}
with a drift $\rho$ (cf. Kwok \cite{Kw}, Wilmott \emph{et al.} \cite{WDH,WHD}).

However, practical analysis of market data shows the need for more realistic models taking into account the aforementioned drawbacks  of the classical Black--Scholes theory. It stimulated development of various nonlinear  option pricing models in which the volatility function is no longer constant, but is a function of the solution $V$ itself. We focus on the case where the volatility depends on the second derivative $\partial^2_S V$ of the option price with respect to the underlying asset price $S$.
\begin{equation} \label{BS}
\partial_ t V+\frac{1}{2}\hat\sigma(S\partial_S^2 V)^2  S^2\partial^2_S V+r S \partial_S V-rV=0,
\end{equation}
where $\hat\sigma(S\partial^2_S V)$ is a function of the product of the asset price and Gamma of the option (Gamma is the second derivative of $V$ with respect to $S$). 

Motivation for studying the nonlinear extensions of the classical Black-Scholes equation (\ref{BS}) with a volatility depending on $S\partial^2_S V$ arises from classical option pricing models taking into account non--trivial transaction costs due to buying and selling assets (cf. Leland \cite{Le}), market feedbacks effects due to large traders choosing given stock--trading strategies (cf. Frey \emph{et al.} \cite{FP,FS}, Sch\"onbucher and Wilmott \cite{SW}), risk from a volatile and unprotected portfolio (see Janda\v{c}ka and \v{S}ev\v{c}ovi\v{c} \cite{JS}) or investors preferences (cf. Barles and Soner \cite{BaSo}). 

One of the first nonlinear models taking into account transaction costs is the Leland model \cite{Le} for pricing the call and put options. This model was further extended by Hoggard, Whalley and Wilmott \cite{HWW} for general type of derivatives. Qualitative and numerical properties of this model were analyzed by Grandits and Schachinger \cite{GS}, Imai \emph{et al.} \cite{Imai2007}, Ishimura \cite{Ishimura2010}, Grossinho and Morais \cite{Gro} and others.

In this model the variance $\hat\sigma^2$ is given by
\begin{equation}\label{sigma_Leland}
\hat\sigma(S\partial^2_S V)^2 = \sigma^2
\left(1-\mathrm{Le \, sgn}\left(S\partial_S^2 V \right)\right)=\left\{ 
\begin{array}{r@ {\quad}l}
    \sigma^2(1-\mathrm{Le}), & \mathrm{if} \, \partial_S^2 V>0, \\
    \sigma^2(1+\mathrm{Le}), & \mathrm{if} \, \partial_S^2 V<0, \\   
\end{array} \right.
\end{equation}
where $\mathrm{Le} =   \sqrt{\frac{2}{\pi}} \frac{C_0}{\sigma \sqrt{\Delta t}}$ is the so-called Leland number, $\sigma$ is a constant historical volatility, $C_0>0$ is a constant transaction costs per unit dollar of transaction in the underlying asset market and $\Delta t$ is the time--lag between consecutive portfolio adjustments. The nonlinear model (\ref{BS}) with the volatility function given as in  (\ref{sigma_Leland}) can be also viewed as a jumping volatility model investigated by Avellaneda and Paras \cite{AP}. 

The important contribution in this direction has been presented in the paper \cite{AAMR} by Amster, Averbuj, Mariani and Rial, where the transaction costs are assumed to be a nonincreasing linear function of the form $C(\xi)=C_0-\kappa\xi$, ($C_0, \,\kappa >0$), depending on the volume of trading stock $\xi\geq0$ needed to hedge the replicating portfolio. A disadvantage of such a transaction costs function is the fact that it may attain negative values when the amount of transactions exceeds the critical value $\xi = C_0/\kappa$. In the model studied by Amster \emph{et al.} \cite{AAMR}  (see also Averbuj \cite{Averbuj2012}, Mariani \emph{et al.} \cite{Mariani2011}) volatility function has the following form:

\begin{equation}\label{sigma:Amster}
\hat\sigma(S\partial^2_S V)^2=\sigma^2 \left( 1 -  \mathrm{Le}\, \mathrm{sgn} \left(S\partial_S^2 V \right) + \kappa  S \partial_S^2 V \right).
\end{equation}

In \cite{BaHowison} Bakstein and Howison investigated a parametrized model for liquidity effects arising from the asset trading. In their model $\hat\sigma$ is a quadratic function of the term $H=S\partial_S^2 V$:
\begin{align}
\hat\sigma(S\partial^2_S V)^2= & \sigma^2 \Biggl(  1+\bar{\gamma}^2(1-\alpha)^2 + 2\lambda  S \partial^2_S V +  \lambda^2(1-\alpha)^2\left(S\partial^2_S V\right)^2  
\nonumber \\ 
&+ 2\sqrt{\frac{2}{\pi}} \bar{\gamma} \, \mathrm{sgn}\left(S\partial^2_S V\right) + 2\sqrt{\frac{2}{\pi}} \lambda (1-\alpha)^2 \bar{\gamma}  \left|S \partial^2_S V \right| \Biggr).
\end{align}
The parameter $\lambda$ corresponds to a market depth measure, i.e. it scales the slope of the average transaction price. Next, the parameter $\bar{\gamma} $ models the relative bid--ask spreads and it is related to the Leland number through relation $2\bar{\gamma}\sqrt{2/\pi}=\mathrm{Le}$. Finally, $\alpha$ transforms the average transaction price into the next quoted price, $0\leq \alpha \leq 1 $.

The risk adjusted pricing methodology (RAPM) model takes into account risk from the unprotected portfolio was proposed by Kratka \cite{Kr}. It was generalized and analyzed by Janda\v{c}ka and \v{S}ev\v{c}ovi\v{c} in \cite{JS}. In this model the volatility function has the form:
\begin{equation} \label{vol_RAPM}
\hat{\sigma}(S\partial^2_S V)^2=\sigma^2
\left(1-\mu\left(S\partial_S^2 V \right)^{\frac{1}{3}} \right),
\end{equation}
where $\sigma>0$ is a constant historical volatility of the asset price return and $\mu=3(C_0^2R/2\pi)^{\frac{1}{3}}$, where $C_0,\,R\geq 0$ are non--negative constants representing cost measure and the risk premium measure, respectively. 

The structure of the paper is as follows. In the next section we present a nonlinear option pricing model under variable transaction cost. It turns out that the volatility function depends on $S\partial_S^2 V$. In particular case of constant or linearly decreasing transaction costs it is a generalization of the Leland \cite{Le} and Amster \emph{et al.} model \cite{AAMR}, respectively. Section 3 is devoted to transformation of the fully nonlinear option pricing equation into a quasilinear Gamma equation. We prove existence of classical H\"older smooth solutions and we derive useful bounds on the solution. In section 4 we propose a numerical scheme for solving the Gamma equation based on finite volume method. We also present several numerical examples of computation of option prices based on a solution to the nonlinear Black--Scholes equation under variable transaction costs.

\section{Option pricing model under variable transaction costs}

One of the key assumptions of the classical Black--Scholes theory is the possibility of continuous adjustment (or hedging) of the portfolio consisting of options and underlying assets. In the context of transaction costs needed for buying and selling the underlying asset, continuous hedging leads to an infinite number of transactions and the unbounded total transaction costs. The Leland  model \cite{Le} (see also Hoggard, Whalley and Wilmott \cite{HWW}) is based on a simple, but very important modification of the Black--Scholes model, which includes transaction costs and possibility of rearranging the portfolio at discrete times. Since the portfolio is maintained at regular intervals, it means that the total transaction costs are limited.

Our derivation of the variable transaction costs option pricing model follows ideas proposed by Leland in \cite{Le}. Therefore we recall crucial steps of derivation of Leland's approach for modeling transaction costs. We assume that the cost $C_0$ per one transaction is constant, i.e. it does not depend on the volume of transactions. The underlying asset is purchased at a higher \textit{ask} price $S_{ask}$ and it is sold for a lower \textit{bid} price $S_{bid}$. The price of $S$ is then computed as an average of \textit{ask} and \textit {bid} prices, i.e., $ S = (S_ {ask} + S_{bid})/2 $. Then $C_0\ge0$ represents a constant percentage of the cost of the sale and purchase of a share relative to the price $S$, i.e. 
\begin{equation}
C_0 = \frac{S_{ask} - S_{bid}}{S} = 2 \, \frac{S_{ask} - S_{bid}}{S_{ask} + S_{bid}}.
\end{equation}

The value $\Pi=  V + \delta S$ of the synthesized portfolio consisting of one option in a long position at the price $V$ and $\delta$ underlying assets at the price $S$ changes over the time interval $[t, t+\Delta t]$ by selling $\Delta \delta<0$ or buying $\Delta \delta>0$ short positioned assets. It means that the purchase or selling $\Delta\delta$ assets at a price of $S$ yields the additional cost $\Delta TC$ for the option holder
\begin{equation}\label{constTC}
\Delta TC = \frac{S}{2} C_0 | \Delta\delta |.
\end{equation}
Consequently, the value of the portfolio changes to:
\begin{equation} \label{P3}
\Delta \Pi = \Delta ( V + \delta S) - \Delta TC
\end{equation}
during the time interval $[t, t+\Delta t]$. The key step in derivation of the Leland model consists in approximation of the change $\Delta TC$ of transaction costs by its expected value $\E[\Delta TC]$, i.e. $\Delta TC \approx  r_{TC} S \Delta t$, 
where the transaction costs measure $r_{TC}$ is defined as the expected value of the change of the transaction costs per unit time interval $\Delta t$ and price $S$:
\begin{equation}
r_{TC}  = \frac{\E[\Delta TC]}{S \Delta t}. 
\end{equation}
Hence equation  (\ref{P3}) describing the change in the portfolio has the form: 
\begin{equation} \label{meanP3}
\Delta \Pi = \Delta ( V + \delta S) - r_{TC} S \Delta t.
\end{equation}
Since the underlying asset follows the geometric Brownian motion we have
\begin{equation}
\Delta S = \rho S \Delta t + \sigma S \Delta W,
\end{equation}
where $\Delta W = W_{t+\Delta t}-W_t$ is the increment of the Wiener process. Now, assuming the change $\Delta \Pi$ in the portfolio is balanced by a bond with the risk-free rate $r\ge 0$, i.e.  $\Delta \Pi= r\Pi \Delta$, using It\=o's lemma for $\Delta V$ and applying the delta  hedging strategy $\delta = - \partial _S V$ we obtain generalization of the Black--Scholes equation
\begin{equation}\label{r_BS_rTC} 
\partial_ t V+\frac{1}{2}\sigma^2 S^2 \partial^2_ S V+r S \partial_ S V-rV -r_{TC} S =0. 
\end{equation}
Furthermore, applying It\={o}'s formula for the function $\delta=-\partial_S V$ we obtain 
\[
\Delta \delta  =  - \sigma S \partial_S^2 V \Delta W = - \sigma S \partial_S^2 V \Phi \sqrt{\Delta t}
\]
plus higher order terms in $\sqrt{\Delta t}$. Here $\Phi \sim N(0,1)$ a normally distributed random variable. Hence, in the lowest order $O(\sqrt{\Delta t})$ we have that
\begin{equation} \label{r:alpha1}
|\Delta \delta| = \alpha |\Phi|, \quad \mathrm{where}\ \ \alpha := \sigma S \left| \partial_S^2 V \right| \sqrt{\Delta t}.
\end{equation}

For the case of constant transaction costs given by (\ref{constTC}), using the fact that $\E[|\Phi|] = \sqrt{2/\pi}$ we obtain
\[
r_{TC} S = \frac{\E[\Delta TC]}{\Delta t} = \frac{1}{2} C_0 S \frac{\E[|\Delta \delta|]}{\Delta t} =  \frac{1}{2} S^2 \sqrt{\frac{2}{\pi}} \frac{C_0}{\sqrt{\Delta t}} \sigma |\partial_S^2 V| 
= \frac{1}{2} \sigma^2 S^2 \mathrm{Le} |\partial_S^2 V|,
\]
where $\mathrm{Le} =   \sqrt{\frac{2}{\pi}} \frac{C_0}{\sigma \sqrt{\Delta t}}$ is the Leland number. Inserting the term $r_{TC} S$ into (\ref{r_BS_rTC}) we obtain the Leland equation (\ref{BS}) with the volatility function given by (\ref{sigma_Leland}).

\medskip

Following derivation of the Leland model we present our approach on modeling variable transaction costs. Large investors can expect a discount due to large amount of transactions. The more they purchase the less they will pay per one traded underlying asset. In general, we will assume that the cost $C$ per one transaction  is a nonincreasing function of the amount of transactions, $|\Delta\delta|$, per  unit of time $\Delta t$, i.e.
\begin{equation} \label{nonlin:TC}
C=C(|\Delta\delta|).
\end{equation}
It means that the purchase of $ \Delta\delta > 0 $ or sales of $ \Delta\delta <$ 0 shares at a price of $ S $, we calculate the additional transaction costs $\Delta TC$ per unit of time $\Delta t$: 
\begin{equation} \label{R_TC}
\Delta TC = \frac{S}{2} C(|\Delta\delta|) |\Delta\delta|.
\end{equation}
Hence the transaction costs measure $r_{TC}$ can be expressed as
\begin{equation}\label{r:def_r_TC}
r_{TC}  = \frac{\E[\Delta TC]}{S\Delta t} = \frac{1}{2}\frac{\E[C(|\Delta\delta|) |\Delta\delta|]}{\Delta t}, 
\end{equation}
where $C$ is the transaction costs function and $\Delta\delta$ is the number of purchased $ \Delta\delta > 0 $ or sold $ \Delta\delta < 0$  shares per unit of time $\Delta t$.

In order to simplify notation we introduce the so-called mean value modification of the transaction costs function $\tilde C(\alpha)$ defined as follows:

\begin{definition}\label{def:tilde_C}
Let $C=C(\xi)$,  $C:\R^+_0 \to \R$, be a transaction costs function. The integral transformation $\tilde{C}:\R^+_0 \to \R$ of the function $C$ defined as follows:
\begin{equation}\label{r:tildeC}
\tilde C(\xi) = \sqrt{\frac{\pi}{2}}\E [C(\xi|\Phi|)|\Phi|] 
= \int_0^\infty C(\xi x) x\, e^{-x^2/2}  dx,
\end{equation}
is called the mean value modification of the transaction costs function. Here $\Phi$ is the random variable with a standardized normal distribution, i.e., $\Phi\sim N(0,1)$.
\end{definition}

Applying definition (\ref{r:tildeC})  to equation (\ref{r:def_r_TC}) we obtain the following expression for the transaction costs measure:
\begin{equation}
r_{TC} = \frac12 \sqrt{\frac{2}{\pi}} \frac{\tilde C(\alpha)\alpha }{\Delta t},
\quad\hbox{where} \quad 
\alpha = \sigma S \left| \partial_S^2 V \right| \sqrt{\Delta t}.
\end{equation}

Inserting the transaction costs measure $r_{TC}$ into (\ref{r_BS_rTC}) we obtain  generalization of the Leland model for the case of arbitrary transaction costs function $C(\xi)$.

\begin{proposition}\label{prop:generalTC}
Let $C:\R^+_0\to \R$ be a measurable and bounded transaction costs function. Then the nonlinear Black--Scholes equation for pricing option under variable transaction costs is given by the nonlinear parabolic equation 
\begin{equation}\label{vtcEq}
\partial_ t V+\frac{1}{2}\hat\sigma(S\partial_S^2 V)^2  S^2\partial^2_S V+r S \partial_S V-rV=0,
\end{equation}
where 
\begin{equation}\label{vtcVol}
\hat\sigma(S\partial^2_S V)^2 = \sigma^2 \left( 
1  - \sqrt{\frac{2}{\pi}} 
\tilde C(\sigma S |\partial_S^2 V| \sqrt{\Delta t}) 
\frac{\mathrm{sgn}( S \partial_S^2 V)}{\sigma\sqrt{\Delta t}}
\right).
\end{equation}
\end{proposition}

In the next two propositions we summarize several useful properties of the mean value modification of a transaction costs function. 

\begin{proposition}\label{prop:C}
Let $C(\xi)$ be a measurable bounded transaction costs function such that $\underline{C}_0 \le C(\xi) \le C_0$ for all $\xi \ge 0$. Then its mean value modification  $\tilde C (\xi)$ is a $C^\infty$ smooth function for $\xi>0$. Furthermore, it has the following properties:
\begin{enumerate}
\item $\tilde C(0) = C(0)$;
\item if $C(+\infty) =\lim_{\xi\to\infty} C(\xi)$ then $\tilde C(+\infty)=C(+\infty)$;
\item $\underline{C}_0 \le \tilde C(\xi) \le C_0 $ for all $\xi\ge 0$;
\item if $C$ is nonincreasing (nondecreasing) then $\tilde C$ is a non- increasing (nondecreasing) function as well;
\item if $C$ is a (non-constant) convex function then $\tilde C$ is a (strictly) convex function.
\end{enumerate}
\end{proposition}

\begin{proposition}\label{prop:C-konvex}
Let $C(\xi)$ be a measurable and bounded transaction costs function which is nonincreasing for $\xi\ge 0$. 
\begin{itemize}
 \item If $C(0)>0$ then  the function $\xi\mapsto \frac{\tilde C(\xi)}{\xi}$ is strictly convex for $\xi>0$. 
 \item If $\underline{C}_0 \le C(\xi) \le C_0$ for all $\xi \ge 0$ then 
 \[
  \tilde C(\xi) + \xi \tilde C^\prime(\xi) \ge 2 \underline{C}_0 -C_0.
 \]
\end{itemize}

\end{proposition}

\begin{figure} 
\begin{center}
\includegraphics[width=0.45\textwidth]{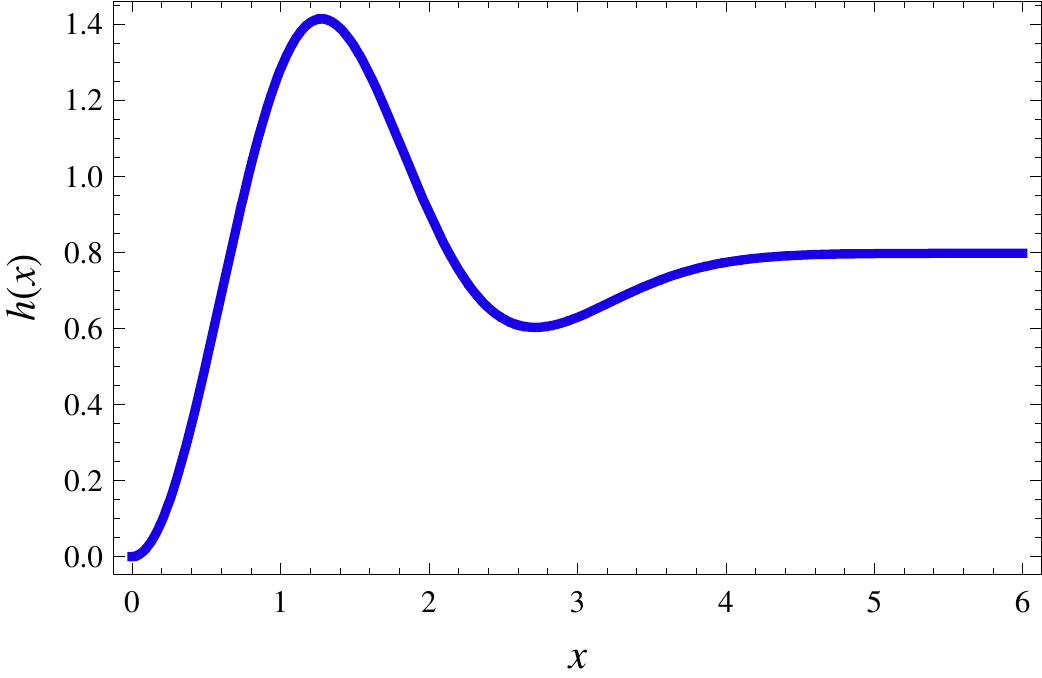}
\caption{Plot of the auxiliary function $h(x)$.}
\label{fig:h-fun}
\end{center}
\end{figure}

\noindent P r o o f. a) Since $\tilde C(\xi) = \sqrt{2\pi} \int_0^\infty C(\xi x) x f(x) dx=\sqrt{2\pi}\xi^{-2} \int_0^\infty C(y) y f(y/\xi) dy$ where $f(x)= e^{-x^2/2}/\sqrt{2\pi}$ and $f^\prime(x) = -x f(x), f^{\prime\prime}(x) = (x^2-1) f(x)$ then, by using substitution $y=\xi x$, we obtain the following identity:
\begin{eqnarray*}
\frac{d^2}{d\xi^2} \left(\frac{\tilde C(\xi)}{\xi} \right)
&=& \sqrt{2\pi} \int_0^\infty C(y) y 
\left( \frac{12}{\xi^5} f(y/\xi) +  \frac{3 y}{\xi^6} f^\prime (y/\xi) +  \frac{y^2}{\xi^7} f^{\prime\prime} (y/\xi) \right) dy \nonumber \\
&=& \frac{\sqrt{2\pi}}{\xi^3} \int_0^\infty C(\xi x) x \left( 12 f(x) + 8 x f^\prime (x) +  x^2  f^{\prime\prime} (x)\right) dx \nonumber \\
&=& \frac{\sqrt{2\pi}}{\xi^3} \int_0^\infty C(\xi x) x \left( x^4 - 9x^2 +12\right) f(x) dx = \frac{\sqrt{2\pi}}{\xi^3} \int_0^\infty C(\xi x) h^\prime(x) dx \nonumber \\
&=& \frac{2 C(+\infty)}{\xi^3} 
-
\frac{\sqrt{2\pi}}{\xi^2} \int_0^\infty C^\prime(\xi x) h(x) dx
>0 \nonumber \\
\end{eqnarray*}
because $C(+\infty) \ge 0, C^\prime(x) \le 0$ (if $C(+\infty)=0$ then $C^\prime \not\equiv 0$) and $h(x)>0$ for $x>0$ where 
\[
h(x)=\int_0^x t(t^4-9t^2+12) f(t) dt = \sqrt{\frac{2}{\pi}} - (2-5x^2 +x^4) f(x) 
\]
is a strictly positive function for $x>0$ depicted in Figure~\ref{fig:h-fun}. It has a unique positive minimum at $x^*=\sqrt{(9+\sqrt{33})/2} \simeq 2.715$.

b) Again, as $\tilde C(\xi) = \xi^{-2}\sqrt{2\pi} \int_0^\infty C(y) y f(y/\xi) dy$, we obtain
\begin{eqnarray*}
\xi \tilde C^\prime(\xi) 
&=& -2 \tilde C(\xi) - \xi^{-3}\sqrt{2\pi} \int_0^\infty C(y) y^2 f^\prime(y/\xi) dy
= -2 \tilde C(\xi) - \sqrt{2\pi} \int_0^\infty C(\xi x) x^2 f^\prime(x) dx
\nonumber \\
&=& -2 \tilde C(\xi) + \sqrt{2\pi} \int_0^\infty C(\xi x) x^3 f (x) dx.
\nonumber \\
\end{eqnarray*}
Since $C(\xi x) \ge \underline{C}_0, \tilde C(\xi) \le C_0$ and $\sqrt{2\pi} \int_0^\infty x^3 f (x) dx=2$ we obtain $\tilde C(\xi) + \xi \tilde C^\prime(\xi)\ge - C_0 + 2 \underline{C}_0$, as claimed. \hfill $\diamondsuit$

\subsection{Piecewise linear nonincreasing transaction costs function}

We present an example of a realistic transaction costs function which is nonincreasing with respect to the amount of transactions as in model studied by Amster \emph{et al.} \cite{AAMR}, Averbuj \cite{Averbuj2012} Mariani \emph{et al.} \cite{Mariani2011}. The benefit is the elimination of the problem of negative values of the linear decreasing costs function. We define the following piecewise linear function.
\begin{definition}\label{def:Cpiecewise}
We define a piecewise linear nonincreasing transaction costs function as follows:
\begin{equation}\label{nonlin:TCPiecewise}
C(\xi) = \left\{ \begin{array}{l@ {\quad} l @{\quad} r}
     C_0,                         &  \mbox{if } &\,     0\le \xi \le \xi_-, \\
     C_0 - \kappa (\xi - \xi_-),  &  \mbox{if } &\, \xi_-\le \xi \le \xi_+, \\
     \underline{C}_0,             &  \mbox{if } &\,          \xi \ge \xi_+, \\
\end{array} \right.
\end{equation}
where we assume $C_0, \, \kappa>0$, and $0\leq\xi_- < \xi_+\leq\infty$ are given constants and $\underline{C}_0=C_0 - \kappa (\xi_+ - \xi_-)>0$. 
\end{definition}
This is a realistic transaction costs function. Indeed, for a small volume of traded assets the transaction costs rate equals $C_0$. When the volume is large enough then a discount is applied with a lower transaction costs rate $\underline{C}_0<C_0$.

Notice that this function also covers several special cases studied before. Namely, the constant transaction costs function studied by Leland \cite{Le}, Hoggard, Whalley and Wilmott \cite{HWW} ($\kappa=\xi_-=0, \xi_+=\infty$) as well as linearly decreasing transaction costs investigated by Amster \emph{et al.} in \cite{AAMR} ($\kappa>0, \xi_-=0, \xi_+=\infty$). 

Taking into account the fact that $C^\prime(\xi) = -\kappa$ for $\xi\in(\xi_-,\xi_+)$ and $C^\prime(\xi)=0$, otherwise, and using integration by parts we can easily derive that the mean value modification of a piecewise transaction costs function (\ref{nonlin:TCPiecewise}) is given by:
\begin{equation} \label{def:tildeCPiecewise}
\tilde C (\xi)=  C_0  - \kappa \xi \int_{\frac{\xi_-}{\xi}}^{\frac{\xi_+}{\xi}} e^{-x^2/2} dx, \quad\hbox{for}\ \xi\ge 0.
\end{equation}

\subsection{Exponentially decreasing transaction costs function}

As an another example of a transaction costs function one can consider the following exponential function of the form
\begin{equation} \label{nonlin:TCExp}
C(\xi) = C_0 \exp(- \kappa \xi), \quad\hbox{for}\ \xi\ge 0,
\end{equation}
where $C_0>0$ and $\kappa>0$ are given constants. Its mean value modification $\tilde C$  can be derived by expanding the function $\xi\mapsto C(\xi|\Phi|)\xi$ into power series:
\begin{eqnarray*}
\tilde C(\xi) &=& \sqrt{\frac{\pi}{2}}\E [C(\xi|\Phi|)|\Phi|] = \int_0^\infty C(\xi x) x\, e^{-x^2/2}  dx = C_0 \int_0^\infty e^{-\kappa \xi x} x\, e^{-x^2/2}  dx 
\\
&=&C_0 \left( \left[ - e^{-\kappa \xi x} e^{-x^2/2}\right]_0^\infty - \int_0^\infty \left(e^{-\kappa \xi x}\right)^\prime  (- e^{-x^2/2})  dx\right)
\\
&=&C_0 \left( 1 - \kappa\xi \int_0^\infty e^{-\kappa \xi x - x^2/2}  dx\right)
=
C_0 \left( 1 - \kappa\xi e^{\kappa^2\xi^2/2} \int_{\kappa\xi}^\infty e^{-t^2/2}  dt\right)
\\
&=& C_0 \phi(-\sqrt{2} \kappa \xi),
\quad\mbox{where}\quad 
\phi(x)=  1 + x e^{\frac{x^2}{4}} \left( \mathrm{erf}(x/2) + 1 \right) \frac{\sqrt{\pi}}{2}, 
\end{eqnarray*}
and $\mathrm{erf}(x/2) = \frac{2}{\sqrt{\pi}} \int_0^{x/2} e^{-s^2} ds$ is the error function.

\begin{figure}
\begin{center}
\includegraphics[width=0.45\textwidth]{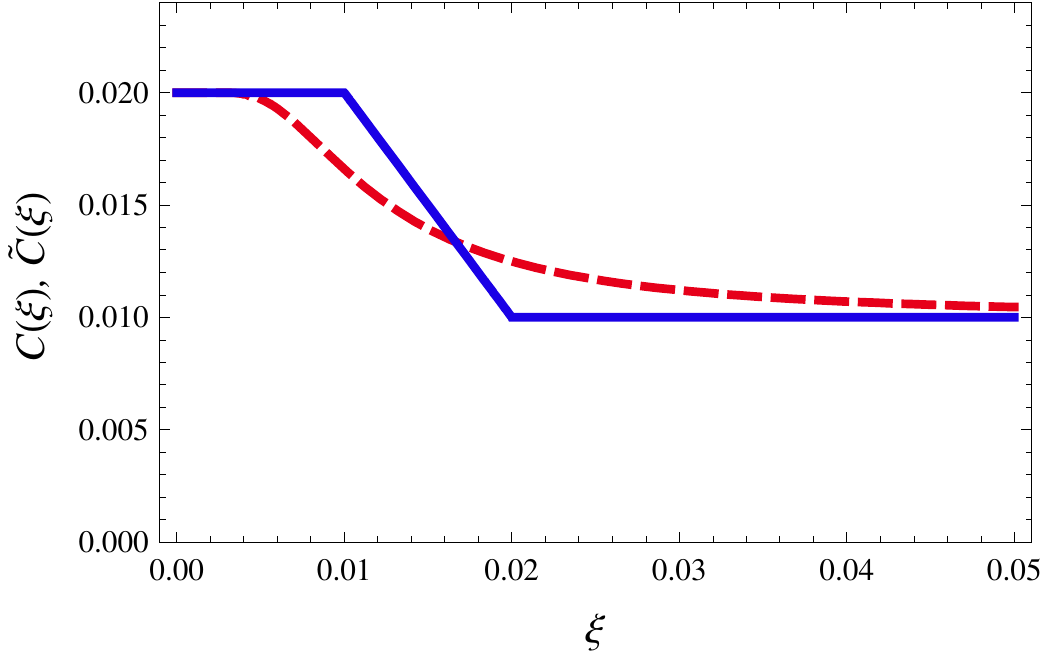}
\includegraphics[width=0.45\textwidth]{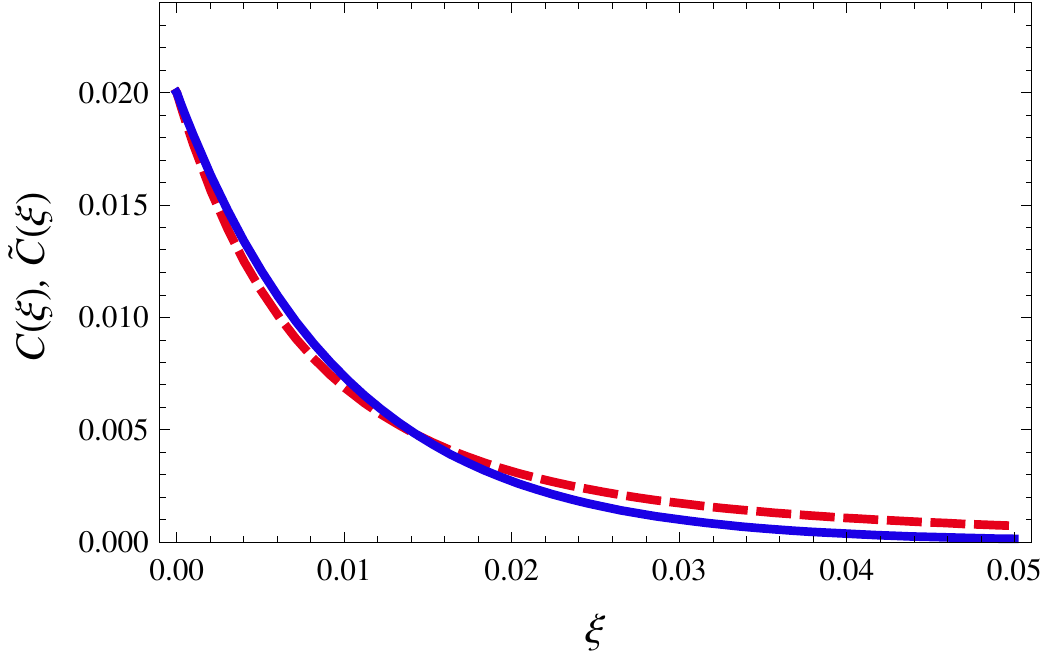}
\\
a) \hskip 6truecm b)
\end{center}

\caption{%
Various types of transaction costs functions $C(\xi)$ (solid line) and their mean value modification $\tilde C(\xi)$ (dashed line). a) Piecewise linear transaction costs function with $C_0 = 0.02, \kappa = 1, \xi_- = 0.01, \xi_+ = 0.02$;
b) Exponentially decreasing transaction costs function with $C_0=0.02, \kappa=100$.
}
\label{fig:Cfun}
\end{figure}

\section{Transformation of the fully nonlinear Black--Scholes equation to the quasilinear Gamma equation}

The goal of this section is to study transformation of the nonlinear Black--Scholes equation into a quasilinear parabolic equation - the so-called Gamma introduced and investigated by Janda\v{c}ka and \v{S}ev\v{c}ovi\v{c} in \cite{JS} (see also \v{S}ev\v{c}ovi\v{c}, Stehl\'{i}kov\'{a} and Mikula \cite[Chapter~11]{SeA}).

In what follows, we will use the notation 
\begin{equation}
 \label{betadef}
 \beta(H) = \frac12 \hat\sigma(H)^2 H,
\end{equation}
where $\hat\sigma$ is the volatility function depending on the term $H=S\partial^2_S V$. Let $E>0$ be a numeraire for the underlying asset price, e.g. $E$ is the expiration price for a call or put option.

\begin{proposition}\label{prop:ekvivalence}
Assume the function $V=V(S,t)$ is a solution to the nonlinear Black-Scholes equation
\begin{equation} \label{r_gamma}
\partial_t V + S \beta (S \partial^2_S V) + r S \partial_S V - rV = 0, \quad S>0, t \in (0,T).
\end{equation} 
Then the transformed function $H=H(x,\tau) = S\partial^2_S V(S,t)$, where $x=\ln(S/E), \tau=T-t$ is a solution to the quasilinear parabolic (Gamma) equation 
\begin{equation} \label{Gamma_rovnica}
\partial_{\tau}H = \partial_x^2 \beta(H) + \partial_x \beta(H)  +  r\partial_x H.
\end{equation}
On the other hand, if $H$ is a solution to (\ref{Gamma_rovnica}) such that $H(-\infty,\tau)=\partial_x H(-\infty,\tau) =0$ and $\beta^\prime(0)$ is finite then the function 
\begin{equation}
 \label{Vexpress}
 V(S,t) = a S + b e^{-r(T-t)} +  \int_{-\infty}^{\infty} (S-Ee^x)^+ H (x,T-t)dx,
\end{equation}
is a solution to the nonlinear Black--Scholes equation (\ref{r_gamma}) for any $a,b\in\R$.
\end{proposition}

\noindent P r o o f.
The first part of the statement can be shown directly by taking the second derivative of (\ref{r_gamma}) with respect to $S$. Indeed, as $\partial_x = S\partial_S$ we have 
\[
S\partial_S^2 (S\beta) = S\partial_S(\beta + S\partial_S\beta) = \partial_x(\partial_x\beta + \beta),
\quad S\partial_S^2(S\partial_S V) = S\partial_S(\partial_S V + S\partial_S^2 V) = H +\partial_x H.
\]
Applying the operator $S\partial_S^2$ to equation (\ref{r_gamma}) and taking into account $\partial_t = -\partial_\tau$ we conclude that $H$ is the solution to equation  (\ref{Gamma_rovnica}) (see also \cite{JS} and \cite[Chapter~11]{SeA}).

On the other hand, if $V(S,t)$ is given by (\ref{Vexpress}) then $S\partial^2_S V(S,t) = H(x,\tau)$. Moreover, if $H$ is a solution to  (\ref{Gamma_rovnica}) then 
\begin{eqnarray*}
\partial_t V(S,t)  &=& r b e^{-r(T-t)} - \int_{-\infty}^{\ln(S/E)} (S-Ee^x) \partial_\tau H (x,\tau)dx \nonumber \\
&=& r b e^{-r(T-t)} - \int_{-\infty}^{\ln(S/E)} (S-Ee^x) 
\left[ e^{-x}\partial_x(e^x \partial_x\beta(H(x,\tau))) + r\partial_x H(x,\tau)\right] dx \nonumber \\
&=& r b e^{-r(T-t)} - \left[ (S-Ee^x)( \partial_x\beta(H(x,\tau))+ r\partial_x H(x,\tau) \right]_{x=-\infty}^{x=\ln(S/E)} 
\nonumber \\
&& - S \int_{-\infty}^{\ln(S/E)} \partial_x\beta(H(x,\tau)) dx - r E \int_{-\infty}^{\ln(S/E)} e^x H(x,\tau) dx
 \nonumber \\
&=& r b e^{-r(T-t)} - S \beta(S\partial^2_S V(S,t))  - r E \int_{-\infty}^{\ln(S/E)} e^x H(x,\tau) dx.
\end{eqnarray*}
Here we have used the fact that $H(-\infty,\tau)=\partial_x H(-\infty,\tau) =0$ and $\beta^\prime(0^\pm)$ is finite. Since 
\[
 S\partial_S V(S,t) = a S + S \partial_S \int_{-\infty}^{\ln(S/E)} (S-Ee^x) H(x,\tau) dx
 = a S + S \int_{-\infty}^{\ln(S/E)}  H(x,\tau) dx
\]
we finally obtain
\[
 \partial_t V + S\beta(S\partial^2_S V) + r S\partial_S V = r(a S + b e^{-r(T-t)}) 
 + r \int_{-\infty}^{\ln(S/E)} (S-Ee^x)  H(x,\tau) dx = r V.
\]
Therefore $V(S,t)$ solves the nonlinear Black--Scholes equation (\ref{r_gamma}), as claimed.
\hfill$\diamondsuit$

\begin{remark}\label{rem:initcond}
If the initial condition $H(x,0)=\delta(x)$ is the Dirac $\delta$-function then for the terminal pay-off diagram $V(S,T)$ given by (\ref{Vexpress}) we obtain 
\begin{enumerate}
\item $V(S,T) = (S-E)^+$ (call option) when $a=b=0$,
\item $V(S,T) = (E-S)^+$ (put option) when $a=-1,b=E$.
\end{enumerate}
\end{remark}

\begin{figure}
\begin{center}
\includegraphics[width=0.42\textwidth]{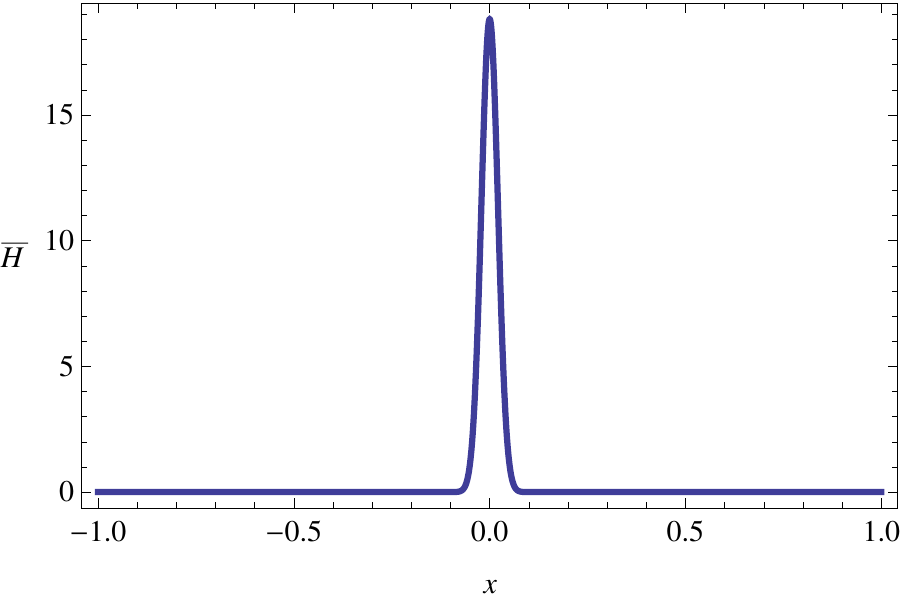}
\quad
\includegraphics[width=0.42\textwidth]{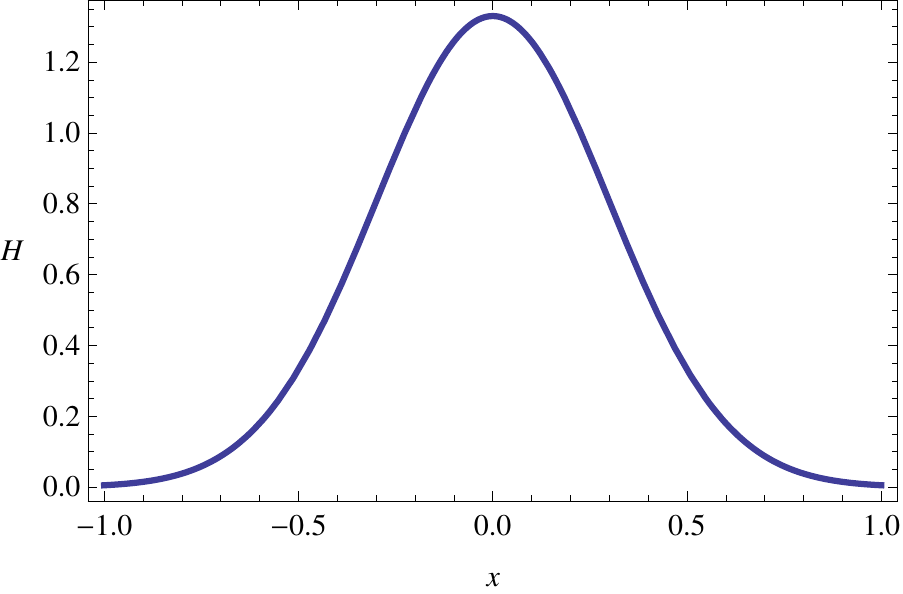}
\medskip
\caption{
Plots of the initial approximation of the function $H(x,0)$ for $\tau^*$ sufficiently small (left) and the solution profile $H(x,T)$ at $\tau=T$ (right). 
}
\label{fig:H_init_end}
\end{center}
\end{figure}

\begin{remark}\label{rem:initcond-approx}
The initial Dirac $\delta$-function can be approximated as follows:
\[
H(x,0) \approx f(d)/(\hat\sigma\sqrt{\tau^*}),
\] 
where $\tau^*>0$ is sufficiently small, $f(d)$ is the PDF function of the normal distribution, $f(d) =e^{-d^2/2}/\sqrt{2\pi}$ and $d=\left(x+ (r-\sigma^2/2)\tau^* \right)/\sigma\sqrt{\tau^*}$ where $\sigma=\hat\sigma(0)$. The idea behind such an approximation follows from observation that for a  solution of the linear Black--Scholes equation with a constant volatility $\sigma>0$ at the time $T-\tau^*$ close to expiry $T$ the value $H(x,\tau^*) = S\partial^2_S V(S, T-\tau^*)$ is given by  $H(x,\tau^*) = f(d)/(\hat\sigma\sqrt{\tau^*})$.  An approximation of the initial condition for $0<\tau^*\ll 1$ is shown in Figure~\ref{fig:H_init_end} (left).
\end{remark}

\subsection{Existence of classical solutions, comparison principle}

The aim of this subsection is to analyze classical smooth solutions to the Cauchy problem for the  quasilinear parabolic equation (\ref{Gamma_rovnica}). Following the methodology based on the so-called Schauder's type of estimates (cf. Ladyzhenskaya \emph{et al.} \cite{LSU}), we shall prove existence and uniqueness of classical solutions to (\ref{Gamma_rovnica}). 

We proceed with a definition of function spaces we will work with. Let $\Omega=(x_L, x_R)\subset\R$ be a bounded interval. We denote $Q_T =\Omega\times (0,T)$ the space-time cylinder. Let $0<\lambda<1$. By $\mathcal{H}^{\lambda}(\Omega)$ we denote the Banach space consisting of all continuous functions $H$ defined on $\bar\Omega$ which are $\lambda$-H\"older continuous. It means that their H\"older semi-norm $\langle H \rangle^{(\lambda)} = \sup_{x,y\in\Omega, x\not= y} |H(x) - H(y)|/|x-y|^\lambda$ is finite. The norm in the space $\mathcal{H}^{\lambda}(\Omega)$ is then the sum of the maximum norm of $H$ and the semi-norm $\langle H \rangle^{(\lambda)}$. The space $\mathcal{H}^{2+\lambda}(\Omega)$ consists of all twice continuously differentiable functions $H$ in $\bar\Omega$ whose second derivative $\partial_x^2 H$ belongs to $\mathcal{H}^{\lambda}(\Omega)$. The space $\mathcal{H}^{2+\lambda}(\R)$ consists of all functions $H:\R\to\R$ such that $H\in \mathcal{H}^{2+\lambda}(\Omega)$ for any bounded domain $\Omega\subset\R$.

The parabolic H\"older space  $\mathcal{H}^{\lambda, \lambda/2}(Q_T)$ of functions defined on a bounded cylinder $Q_T$ consists of all continuous functions $H(x,\tau)$ in $\bar{Q}_T$  such that $H$ is $\lambda$-H\"older continuous in the $x$-variable and $\lambda/2$-H\"older continuous in the $t$-variable. The norm is defined as the sum of the maximum norm and corresponding H\"older semi-norms. The space $\mathcal{H}^{2+\lambda, 1+\lambda/2}(Q_T)$ consists of all continuous functions on $\bar{Q}_T$ such that $\partial_\tau H, \partial^2_x H \in \mathcal{H}^{\lambda, \lambda/2}(Q_T)$. Finally, the space $\mathcal{H}^{2+\lambda, 1+\lambda/2}(\R\times [0,T])$ consists of all functions $H:\R\times [0,T]\to \R$ such that $H \in \mathcal{H}^{2+\lambda, 1+\lambda/2}(Q_T)$ for any bounded cylinder $Q_T$ (cf. \cite[Chapter I]{LSU}).

We first derive useful lower and upper bounds of a solution $H$ to the Cauchy problem (\ref{Gamma_rovnica}). The idea of proving upper and lower estimates for $H(x,\tau)$ is based on construction of suitable sub- and super-solutions to the parabolic equation (\ref{Gamma_rovnica}) (cf. \cite{LSU}).

\begin{lemma}
\label{th:CompPsi} 
Suppose that the initial condition $H(.,0)\in \mathcal{H}^\lambda(\R)$ is non-negative and uniformly bounded from above, i.e., $\overline{H} =\sup_{x\in\R} H(x,0) <\infty$. Assume $\beta(H)$ is a $C^{1,\varepsilon}$ smooth function for $H\ge 0$ and satisfying the following strong parabolic inequalities:
\[
\lambda_- \le \beta^\prime(H) \le \lambda_+ 
\]
for any $H\ge 0$ where $\lambda_\pm>0$ are constants. If the bounded solution $H(x,\tau)$ to the  quasilinear parabolic equation (\ref{Gamma_rovnica}) belongs to the function space $\mathcal{H}^{2+\lambda, 1+\lambda/2}(\R\times [0,T])\cap L_\infty(\R\times(0,T))$, for some $0<\lambda<1$, then it satisfies the following inequalities:
\[
0 \le  H(x,\tau) \le  \overline{H}, \quad \hbox{for any}\ \tau\in [0,T)\ \hbox{and}\ x\in\R.
\]
\end{lemma}

\noindent P r o o f.
The quasilinear parabolic equation  (\ref{Gamma_rovnica}) can be rewritten in the form:
\begin{equation} \label{fullynonlinear}
\partial_{\tau}H = \partial_x( \beta^\prime(H) \partial_x H) + \beta^\prime(H) \partial_x H  +  r\partial_x H.
\end{equation}
Notice that the right-hand side of (\ref{fullynonlinear}) is a strictly parabolic operator because $0< \lambda_- \le \beta^\prime(H) \le \lambda_+$. Since the constant functions $\underline{H} \equiv 0$ and $\overline{H}$ are solutions to (\ref{fullynonlinear}) then the statement is a consequence of  the parabolic comparison principle for strongly
parabolic equations (see e.g. \cite[Chapter V, (8.2)]{LSU}). \hfill$\diamondsuit$

\medskip
In the next Proposition we show that the diffusion function $\beta(H)$ corresponding to the nonlinear Black--Scholes equation for pricing options under the variable transaction costs satisfies the strong parabolicity assumption. 

\begin{proposition}\label{betaderprop}
Let $C(\xi)$ be a measurable bounded transaction costs function which is nonincreasing 
and such that $\underline{C}_0 \le C(\xi) \le C_0$ for all $\xi \ge 0$. Let 
\[
\hat\sigma (H)^2 = \frac{\sigma^2}{2} 
\left(1 - \sqrt{\frac{2}{\pi}} \frac{\tilde C(\sigma\sqrt{\Delta t} |H|)}{\sigma \sqrt{\Delta t}} \hbox{sgn}(H)\right).
\]
Then for the diffusion function $\beta(H)=\hat\sigma(H)^2 H$ the following inequalities hold:
\[
\frac{\sigma^2}{2}(1-\mbox{Le}) \le \beta^\prime(H) \le  \frac{\sigma^2}{2}(1-2\mbox{\underline{Le}} + \mbox{Le}) 
\]
for all $H\ge0$, where $\hbox{\underline{Le}} =  \sqrt{\frac{2}{\pi}} \frac{\underline{C}_0}{\sigma\sqrt{\Delta t}}$ and $\hbox{Le} =  \sqrt{\frac{2}{\pi}} \frac{C_0}{\sigma\sqrt{\Delta t}}$.
\end{proposition}

\noindent P r o o f. 
For $H\ge 0$ we have
\[
\frac{2}{\sigma^2} \beta^\prime(H) =  
1-  \sqrt{\frac{2}{\pi}} \frac{1}{\sigma\sqrt{\Delta t}}
\left(
\tilde C(\xi) +\xi \tilde C^\prime(\xi)\right), \quad \mbox{where}\ \ 
\xi\equiv\sigma\sqrt{\Delta t} H. 
\]
Since $C$ is a nonincreasing function then $\tilde C^\prime(\xi)\le 0$ and 
$\tilde C(\xi)\le C_0$ (see Proposition~\ref{prop:C}) then the inequality $\frac{\sigma^2}{2}(1-\mbox{Le}) \le \beta^\prime(H)$ easily follows. According to Proposition~\ref{prop:C-konvex} we have $\tilde C(\xi) +\xi \tilde C^\prime(\xi) \ge 2\underline{C}_0 -C_0$ and so $\beta^\prime(H)\le \frac{\sigma^2}{2}(1-2 \mbox{\underline{Le}} + \mbox{Le})$ and the proof of the statement follows. \hfill$\diamondsuit$

\begin{theorem}\label{existence}
Suppose that the initial condition $H(.,0)\ge 0$ belongs to the H\"older space $\mathcal{H}^{2+\lambda}(\R)$ for some $0<\lambda< \min(1/2,\varepsilon)$ and  $\overline{H} =\sup_{x\in\R} H(x,0) <\infty$. Assume that $\beta\in C^{1,\varepsilon}$ satisfies $\lambda_- \le \beta^\prime(H) \le \lambda_+$ for any $0\le H\le \overline{H}$ where $\lambda_\pm>0$ are constants.

Then there exists a unique classical solution $H(x,\tau)$ to the quasilinear parabolic equation (\ref{Gamma_rovnica}) satisfying the initial condition  $H(x,0)$. The function $\tau\mapsto \partial_\tau H(x,\tau)$ is $\lambda/2$-H\"older continuous for all $x\in\R$ whereas $x\mapsto\partial_x H(x,\tau)$ is Lipschitz continuous for all $\tau\in[0,T]$. Moreover, $\beta(H(.,.))\in \mathcal{H}^{2+\lambda, 1+\lambda/2}(\R\times [0,T])$ and $0<H(x,\tau) \le \overline{H}$ for all $(x,\tau)\in\R\times[0,T)$.
\end{theorem}

\noindent P r o o f. 
The proof is based on the so-called Schauder's theory on existence and uniqueness of classical H\"older smooth solutions to a quasi-linear parabolic equation of the form (\ref{Gamma_rovnica}). It follows the same ideas as the proof of \cite[Theorem 5.3]{KS} where Kilianov\'a and \v{S}ev\v{c}ovi\v{c} investigated a similar quasilinear parabolic equation obtained from a nonlinear Hamilton-Jacobi-Bellman equation in which a stronger assumption $\beta\in C^{1,1}$ is assumed. Nevertheless, we sketch the key steps of the proof.

The Schauder theory requires that the diffusion coefficient of a quasi-linear parabolic equation is sufficiently smooth. Therefore the function $\beta$ has to be regularized by a $\delta$-parameterized family of smooth mollifier functions $\beta_{(\delta)}(H)$ such that $\beta_{(\delta)} \rightrightarrows \beta$, and  
$\beta^\prime_{(\delta)} \rightrightarrows \beta^\prime$ locally uniformly as $\delta\to 0$. For any $\delta>0$, 
the existence of the unique classical bounded solution $H^\delta\in\mathcal{H}^{2+\lambda, 1+\lambda/2}(\R\times [0,T])\cap L_\infty(\R\times(0,T))$ to the Cauchy problem:
\begin{equation*}
\partial_\tau H^\delta - \partial_{x} (\beta^\prime_{(\delta)}(H^\delta) \partial_x H^\delta ) 
= \partial_x f(H^\delta, \beta_{(\delta)}(H^\delta)), 
 \quad H^\delta(x,0) = H(x,0), \quad x\in\R, t\in[0,T),
\end{equation*}
follows from \cite[Theor. 8.1 and Rem. 8.2, Ch. V, pp. 495--496]{LSU}. Here $f(H, \beta(H)) := \beta(H) + r H$. 

By virtue of Lemma~\ref{th:CompPsi}, $H^\delta, 0<\delta\ll 1$, is uniformly bounded in the space $L_\infty(Q_T)$ for any bounded cylinder $Q_T$. Using the inequality \cite[Chapter I, (6.6)]{LSU} we can prove that $H^\delta, 0<\delta\ll 1$, is also uniformly bounded in the Sobolev space  $W^1_2(Q_T)$. It means that there exists a subsequence $H^{\delta_k}\rightharpoonup H$ weakly converging to a function $H\in W^1_2(Q_T)$ as $\delta_k\to 0$. As a consequence of the Rellich-Kondrashov compactness embedding theorem  $W^1_2(Q_T) \hookrightarrow L_2(Q_T)$ (cf. \cite[Chapter II, Theorem 2.1]{LSU}) the limiting function $H\in W^1_2(Q_T)$ is a weak solution to the quasi-linear parabolic equation (\ref{Gamma_rovnica}). Since
$H, f\in W^1_2(Q_T)$ we obtain $\partial^2_x \beta(H) \in L_2(Q_T)$. Furthermore, $\partial_\tau \beta(H) \in L_2(Q_T)$ because $\lambda_-< \beta^\prime(H)<\lambda_+$. Hence, $\beta(H)$ belongs to the parabolic Sobolev space $W^{2,1}_2(Q_T)$ which is continuously embedded into the H\"older space 
$\mathcal{H}^{\lambda, \lambda/2}(Q_T)$ for any $0<\lambda<\min(1/2,\varepsilon)$. 

Finally, the transformed function $z(x,\tau) := \beta(H(x,\tau))$ is a solution to the quasi-linear parabolic equation in the non-divergent form: 
\[
\partial_\tau z =  \zeta(z) \left[ \partial^2_x z  + \partial_x f(\alpha(z), z)
\right] =0, \qquad z(x,0) = \beta(H(x,0)),
\]
where $\zeta(z) = \beta^\prime(\alpha(z))$ and $z\mapsto \alpha(z)$ is the inverse function to the increasing function $H\mapsto \beta(H)$. The function $z\mapsto \beta^\prime(z)$ is $\varepsilon$-H\"older continuous. Thus  $z\mapsto \zeta(z)$ is $\varepsilon-$ H\"older continuous as well. Now, we can apply a simple boot-strap argument to show that $z=z(x,\tau)$ is sufficiently smooth. Clearly, it is a solution to the linear parabolic equation in non-divergence form
\[
\partial_\tau z = a(x,\tau) \partial^2_x z  + b(x,\tau) \partial_x z, \qquad z(x,0) = \beta(H(x,0)),
\]
where $a(x,\tau):=\zeta(z(x,\tau)), b(x,\tau) = \zeta(z(x,\tau)) \left( 1 + r \alpha^\prime (z(x,\tau)) \right)$. The functions $a$ and $b$ belong to the H\"older space $H^{\lambda,\lambda/2}(Q_T)$ because $z\in H^{\lambda, \lambda/2}(Q_T)$. With regard to \cite[Theorem 12.2, Chapter III]{LSU} we have $z=\beta(H)\in H^{2+\lambda, 1+\lambda/2}(Q_T)$ and the proof now follows
because the domain $Q_T\subset \R\times (0,T)$ was arbitrary. \hfill$\diamondsuit$

\section{Numerical full space-time discretization scheme for solving the Gamma equation}
\label{sec5:fdm}

The purpose of this section is to derive an efficient numerical scheme for solving the Gamma equation. The construction of numerical approximation of a solution $H$ to (\ref{Gamma_rovnica}) is based on a derivation of a system of difference equations corresponding to (\ref{Gamma_rovnica}) to be solved at every discrete time step. We make use of the numerical scheme adopted from the paper by Janda\v cka and \v Sev\v covi\v c \cite{JS} in order to solve the Gamma  equation (\ref{Gamma_rovnica}) for a general function $\beta=\beta(H)$ including, in particular, the case of the model with variable transaction costs.  The efficient numerical discretization is based on the finite volume approximation of the partial derivatives entering (\ref{Gamma_rovnica}). The resulting scheme is semi--implicit in a finite--time difference approximation scheme. 

Other finite difference numerical approximation schemes are based on discretization of the original fully nonlinear Black--Scholes equation in non-divergence form (\ref{BS}). We refer the reader to recent publications by Ankudinova and Ehrhardt \cite{AE}, Company \emph{et al.} \cite{CompanyNavaro}, D\"uring \emph{et al.} \cite{DFJ}, Liao and Khaliq \cite{LiaK}, Zhou \emph{et al.} \cite{Zhou2015}. Recently, a quasilinearization technique for solving the fully nonlinear parabolic equation (\ref{BS}) was proposed and analyzed by Koleva and Vulkov \cite{Koleva}. Our approach is based on a solution to the quasilinear Gamma equation written in the divergence form, so we can use existing finite volume based numerical scheme to solve the problem efficiently (cf. Janda\v{c}ka and \v{S}ev\v{c}ovi\v{c} \cite{JS}, K\'utik and Mikula \cite{KutikMikula}).

For numerical reasons we restrict the spatial interval to $x\in(-L,L)$ where $L>0$ is sufficiently large. Since $S=E e^x \in (E e^{-L}, E e^L)$ it is sufficient to take $L\approx 2$ in order to include the important range of values of $S$. For the purpose of construction of a numerical scheme, the time interval $[0,T]$ is uniformly divided with a time step $k=T / m$ into discrete points $\tau_j= jk$, where $j=0,1, \cdots, m$. We consider the spatial interval $[-L,L]$ with uniform division with a step  $h=L/n$, into discrete points $x_i = ih,$ where$\ i=-n,\cdots,n$.

The proposed numerical scheme is semi--implicit in time. Notice that the term $\partial_x^2\beta,$  can be expressed in the form $\partial_x^2\beta = \partial_x\left( \beta^\prime(H) \partial_x H \right)$, where $\beta^\prime$ is the derivative of  $\beta(H)$ with respect to $H$. In the discretization scheme,  the nonlinear terms $\beta^\prime(H)$ are evaluated from the previous time step $\tau_{j-1}$ whereas linear terms are solved at the current time level.

Such a discretization scheme leads to a  solution of a tridiagonal system of linear equations at every discrete time level.  First, we replace the time derivative by the time difference, approximate $H$ in nodal points by the average value of neighboring segments, then we collect all linear terms at the new time level $\tau_j$ and by taking all the remaining terms from the previous time level $\tau_{j-1}$. We obtain a tridiagonal  system for the solution vector 
$H^j=(H^j_{-n+1}, \cdots, H^j_{n-1})^T \in \R^{2n -1}$:
\begin{equation} \label{doplnky-jam-tridiagonal-bs}
a_{i}^j H_{i-1}^j+b_{i}^j H_{i}^j+c_{i}^j H_{i+1}^j = d_i^j,
\quad H_{-n}^j= 0,\ \ H_n^j =0 \,,
\end{equation}
where $i=-n+1,\cdots,n-1$ and $j=1, \cdots, m$. The coefficients of the tridiagonal matrix are given~by
\[
a_i^j = -\frac{k}{h^2}\beta^\prime_H(H_{i-1}^{j-1}) + \frac{k}{2h}r\,\quad
c_i^j = -\frac{k}{h^2}\beta^\prime_H(H_{i}^{j-1}) - \frac{k}{2h}r\,, \quad 
b_i^j = 1 - (a_i^j + c_i^j)\,,
\]
\[
d_i^j = H_{i}^{j-1} + \frac{k}{h}\Big(\beta(H_{i}^{j-1}) - \beta(H_{i-1}^{j-1} )\Big)\,.
\]
It means that the  vector $H^j$ at the time level $\tau_j$ is a solution to the system of linear equations ${\bf A}^{(j)}\, H^j = d^{j},$ where the $(2n-1)\times (2n-1)$ matrix ${\bf A}^{(j)}=\mbox{tridiag}(a^j,b^j,c^j)$. In order to solve the tridiagonal system in every time step in a fast and  effective way, we can use the efficient Thomas algorithm.

Finally, with regard to Proposition~\ref{prop:ekvivalence} and Remark~\ref{rem:initcond} the option price  $V(S,T - \tau_j)$ can be constructed from the discrete solution $H^j_i$ by means of a simple integration scheme:
\begin{eqnarray}
\hbox{(call option)}\qquad\qquad V(S,T - \tau_j) &=&  h \sum_{i=-n}^n (S- E e^{x_i} )^+ H_i^j, \quad j=1, \cdots, m,
\nonumber \\
\hbox{(put option)}\qquad\qquad V(S,T - \tau_j) &=&  h \sum_{i=-n}^n (E e^{x_i} -S )^+ H_i^j, \quad j=1, \cdots, m.
\nonumber
\end{eqnarray}

\subsection{Numerical results for the nonlinear model with variable transaction costs}

In this section we present the numerical results for computation of the option price.  As an example for numerical approximation of a solution we consider the nonlinear Black--Scholes equation for pricing options under variable transaction costs described by the piecewise linear nonincreasing function, depicted in Figure~\ref{fig:Cfun}. The function $\beta(H)$ corresponding to the variable transaction costs function $C(\xi)$ has the form 
\begin{equation*}
\beta(H) = \frac{\sigma^2}{2} \left( 1 - \tilde{C}(\sigma|H|\sqrt{\Delta t}) \frac{\mathrm{sgn}(H)}{\sigma\sqrt{\Delta t}} \right) H,
\end{equation*}
where $\tilde C$ is the modified transaction costs function.  We assume that the hedging time $\Delta t$ is such that $\mbox{Le} <1$. 

\begin{figure}
\begin{center}
\includegraphics[width=0.45\textwidth]{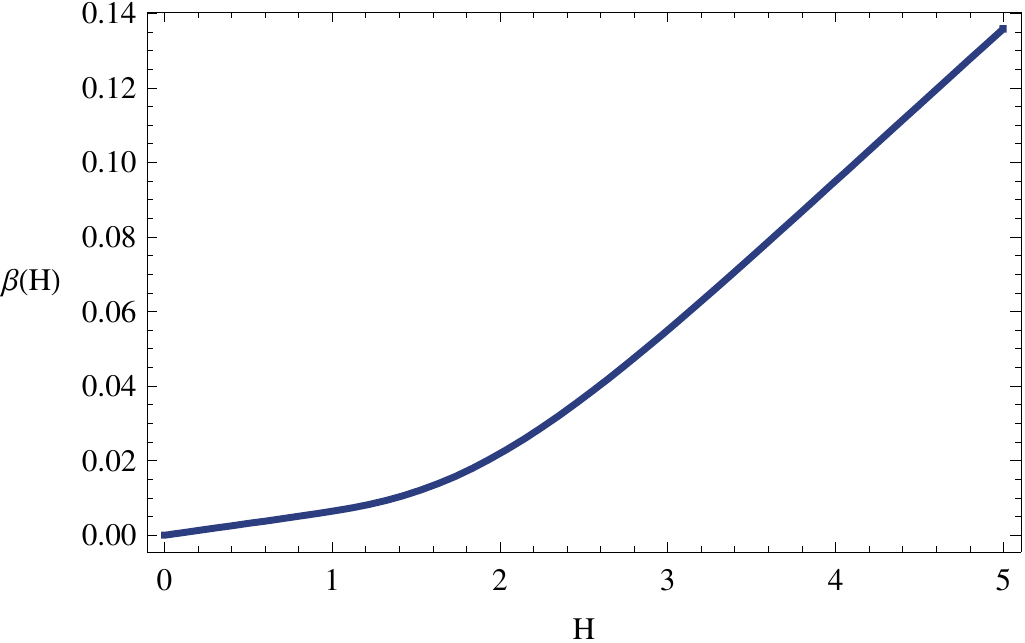}

\caption{
A graph of the function $\beta(H)$ corresponding to the nonlinear model with a piecewise linear variable transaction costs function. 
}
\label{fig:nonlinearTC-betafun}
\end{center}
\end{figure}

\begin{figure} 
\begin{center}
\includegraphics[width=0.48\textwidth]{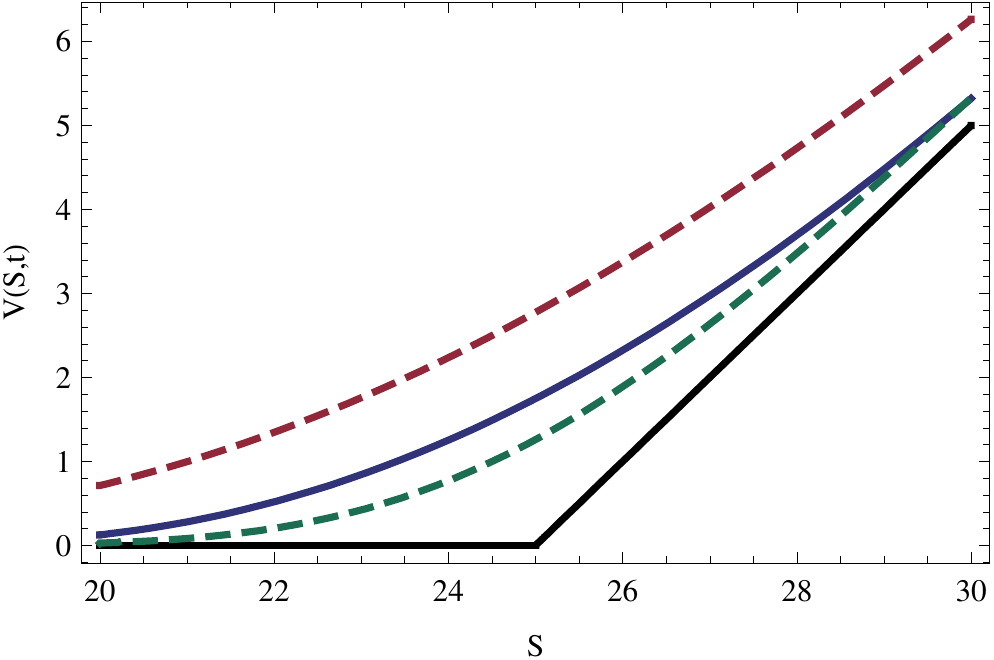}
\quad
\includegraphics[width=0.48\textwidth]{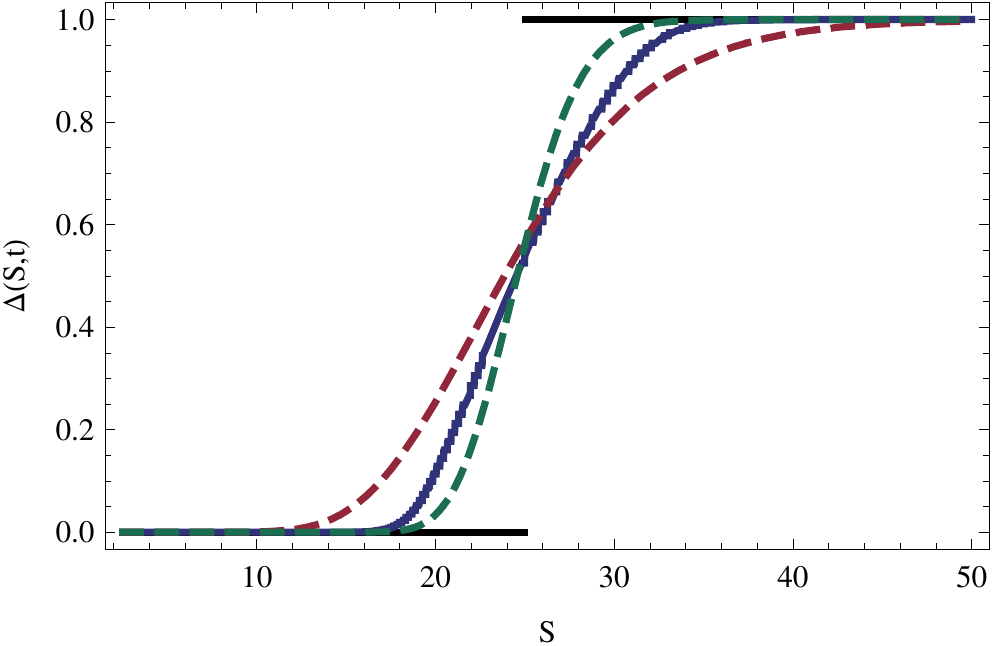}
\centerline{$t=0$}
\medskip

\includegraphics[width=0.48\textwidth]{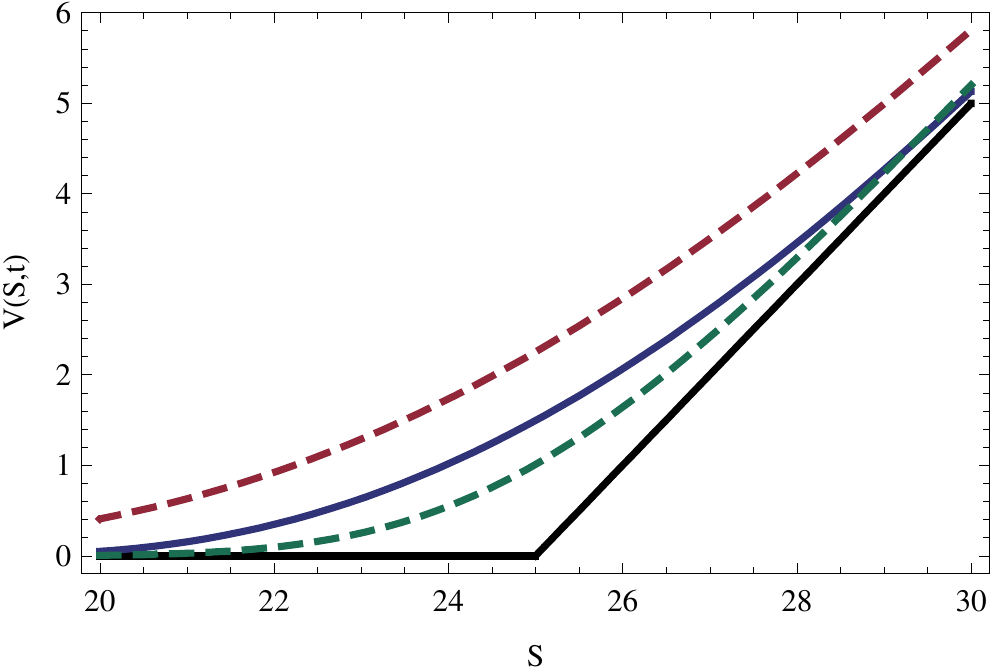}
\quad
\includegraphics[width=0.48\textwidth]{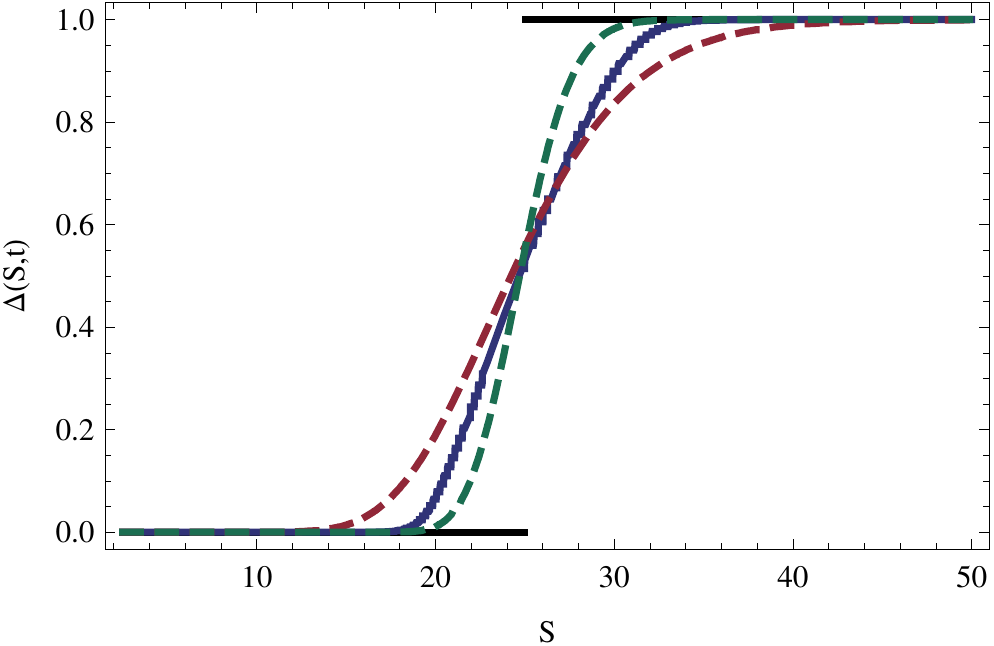}
\centerline{$t=T/3$}
\medskip

\includegraphics[width=0.48\textwidth]{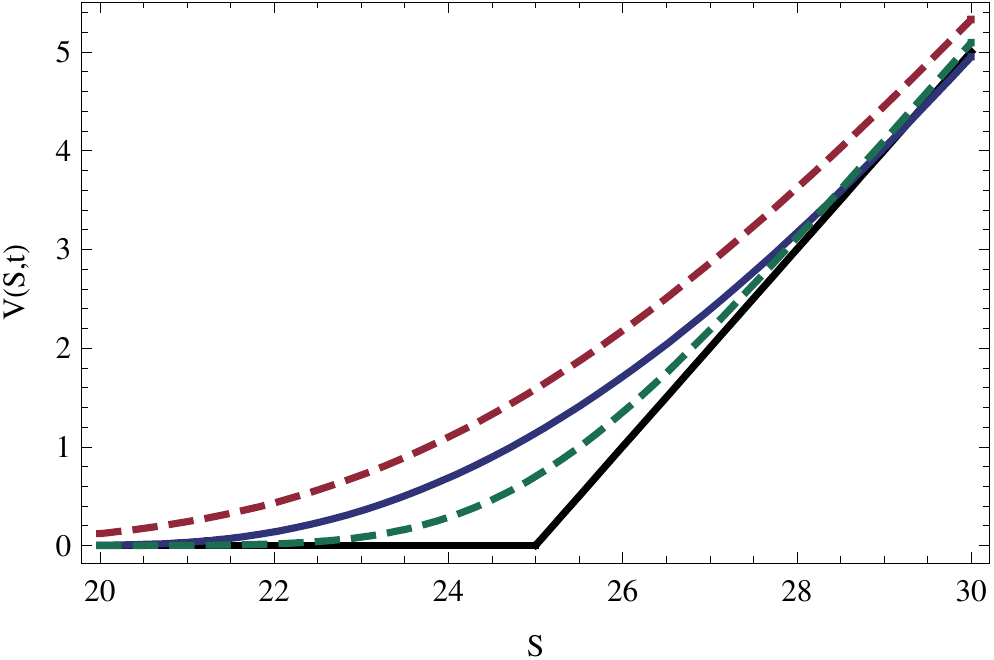}
\quad
\includegraphics[width=0.48\textwidth]{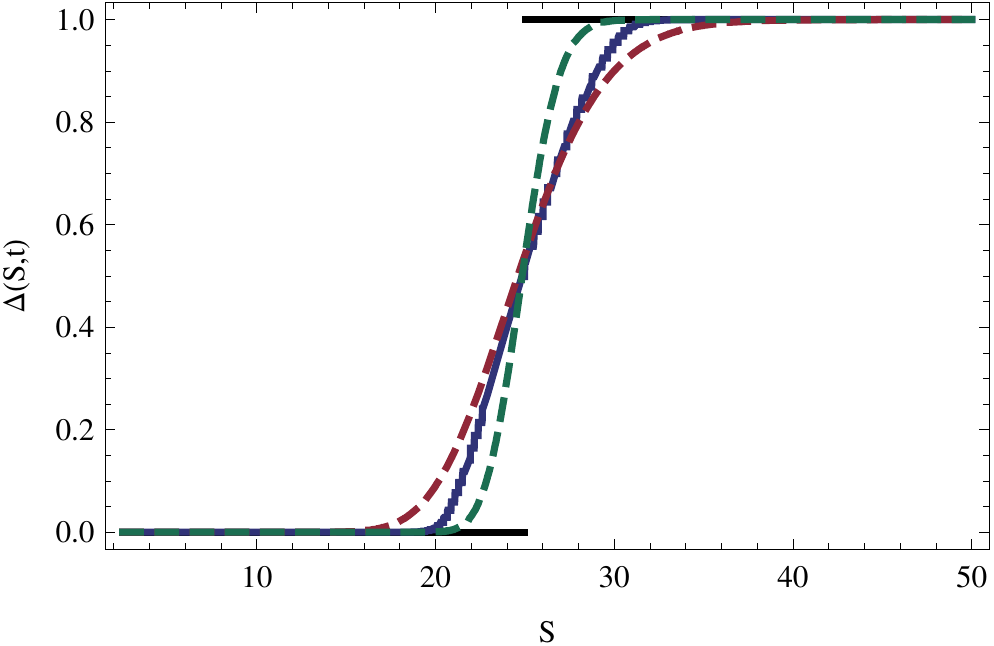}
\centerline{$t=2T/3$}
\medskip

\caption{
The call option price $V(S,t)$ as a function of $S$ for $t\in\{0,T/3, 2T/3\}$ (left) and its delta  $\Delta(S,t)=\partial_S V(S,t)$.
}
\label{fig:nonlinearTC-SolutionDelta}
\end{center}
\end{figure}

In our computations we chose the following model parameters describing the piecewise transaction costs function: $C_0 = 0.02,  \kappa = 0.3, \xi_- = 0.05, \xi_+ = 0.1$. The length of the time interval between two consecutive portfolio rearrangements: $\Delta t = 1/261$. The maturity time $T = 1$, historical volatility $\sigma = 0.3$ and the risk-free interest rate $r = 0.011$. As for the numerical parameters we chose $L=2.5, n = 250, m = 200$, and the parameter $\tau^* = 0.005$ for approximation of the initial Dirac $\delta$--function. The parameters $C_0, \sigma, \kappa, \xi_\pm$ and $\Delta t$ correspond to the Leland numbers $\mbox{Le}=0.85935$ and $\mbox{\underline{Le}}=0.21484$. In Table~\ref{tab:bid_price} we present option values $V_{vtc}(S,0)$ for different prices of the underlying asset achieved by a numerical solution. In Figure~\ref{fig:nonlinearTC-SolutionDelta} we plot the solution $V_{vtc}(S,t)$ and the option price delta factor $\Delta(S,t)=\partial_S V (S,t)$, for various times $t \in \{0,T/3, 2T/3\}$. The upper dashed line corresponds to the solution of the linear Black--Scholes equation with the higher volatility $\hat \sigma^2_{max}=\sigma^2\left(1-\underline{C}_0\sqrt{\frac{2}{\pi}}\frac{1}{\sigma\sqrt{\Delta t}}\right)$, where $\underline{C}_0=C_0 - \kappa (\xi_+ - \xi_-)>0$, whereas the lower dashed line corresponds to the solution with a lower volatility $\hat \sigma^2_{min}=\sigma^2\left(1-{C_0}\sqrt{\frac{2}{\pi}}\frac{1}{\sigma\sqrt{\Delta t}}\right)$.

\begin{table}
\caption{Call option values computed by means of the numerical solution of nonlinear Black--Scholes model in comparison to solutions $V_{\sigma_{max}}, V_{\sigma_{min}}$ computed by the linear Black--Scholes equation with constant volatility $\sigma=\sigma_{max}$ and $\sigma=\sigma_{min}$.}
\begin{center}
{\small
\begin{tabular}{c|ccc}
\hline  $S$  & $V_{\sigma_{max}}(S,0) $        &   $V_{vtc}(S,0)$    &    $ V_{\sigma_{min}}(S,0) $        \\ 
\hline
  20 & 0.709   &   0.127   &   0.029   \\
  23 & 1.752   &   0.844   &   0.421    \\
  25 & 2.768   &   1.748   &   1.258     \\
  28 & 4.723   &   3.695   &   3.474     \\
  30 & 6.256   &   5.321   &   5.327     \\
\hline 
\end{tabular}}
\end{center}
\label{tab:bid_price}
\end{table}

The empirically observed fact (see Table~~\ref{tab:bid_price}) that 
\[
V_{\sigma_{min}}(S,t) \leq V_{vtc} (S,t) \leq V_{\sigma_{max}}(S,t) \quad \hbox{for all}\  S>0, t\in[0,T], 
\]
can be proved analytically. It is a consequence of the parabolic comparison principle (cf. \cite[Chapter V, (8.2)]{LSU}). Indeed, the fully nonlinear Black--Scholes equation  (\ref{r_gamma}) can be considered as a nonlinear strongly parabolic equation 
\begin{equation}\label{abstrakteq}
\partial_\tau V = {\mathcal F}(S, V, \partial_S V, \partial^2_S V),
\end{equation}
for the option price $V=V(S, T-\tau)$, where
\[
{\mathcal F}(S, V, \partial_S V, \partial^2_S V) = 
\frac{1}{2} \sigma^2 
\left( 1  - \sqrt{\frac{2}{\pi}} \tilde C(\sigma S |\partial_S^2 V| \sqrt{\Delta t}) 
\frac{\mathrm{sgn}( S \partial_S^2 V)}{\sigma\sqrt{\Delta t}}
\right) S^2 \partial_S^2 V + r S \partial_S V-rV.
\]
Recall that $0<\lambda_-\le \partial_Q {\mathcal F}(S, V, P, Q) \le \lambda_+$ for all $S,V,P$ and $Q\ge0$ and so equation (\ref{abstrakteq}) is indeed a strongly parabolic equation. For the solution $V_{\sigma_{min}}(S,t)$ of the linear Black--Scholes equation with the constant volatility $\sigma^2_{min}= \sigma^2(1-\mbox{Le})$ we have
\[
\partial_\tau V_{\sigma_{min}} = \frac{1}{2} \sigma^2 ( 1  - \mbox{Le}) S^2 \partial_S^2 V_{\sigma_{min}} + r S \partial_S V_{\sigma_{min}} -r V_{\sigma_{min}} 
\le {\mathcal F}(S, V_{\sigma_{min}}, \partial_S V_{\sigma_{min}}, \partial^2_S V_{\sigma_{min}}),
\]
because $\tilde C(\xi)\le C_0$ and so 
\[
\sqrt{\frac{2}{\pi}} \tilde C(\sigma S |\partial_S^2 V_{\sigma_{min}} | \sqrt{\Delta t}) \frac{1}{\sigma\sqrt{\Delta t}} \le \sqrt{\frac{2}{\pi}} C_0 \frac{1}{\sigma\sqrt{\Delta t}}
= \mbox{Le}.
\]
Hence $V_{\sigma_{min}}$ is a sub-solution to the strongly parabolic equation (\ref{abstrakteq}). Therefore, by the parabolic comparison principle,  $V_{\sigma_{min}}(S,t) \le V_{vtc} (S,t)$ for all $S>0$ and $t\in[0,T]$. Analogously, the inequality $V_{vtc} (S,t) \leq V_{\sigma_{max}}(S,t)$ follows from the parabolic comparison principle because $\tilde C(\xi)\ge \underline{C}_0$.

\begin{figure}
\begin{center}
\includegraphics[width=0.48\textwidth]{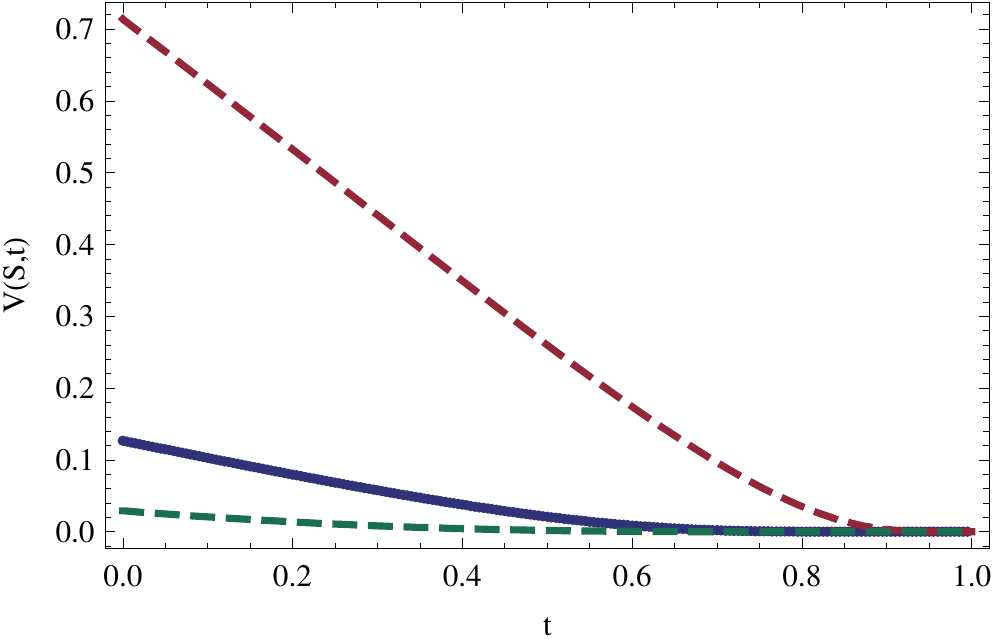}
\includegraphics[width=0.48\textwidth]{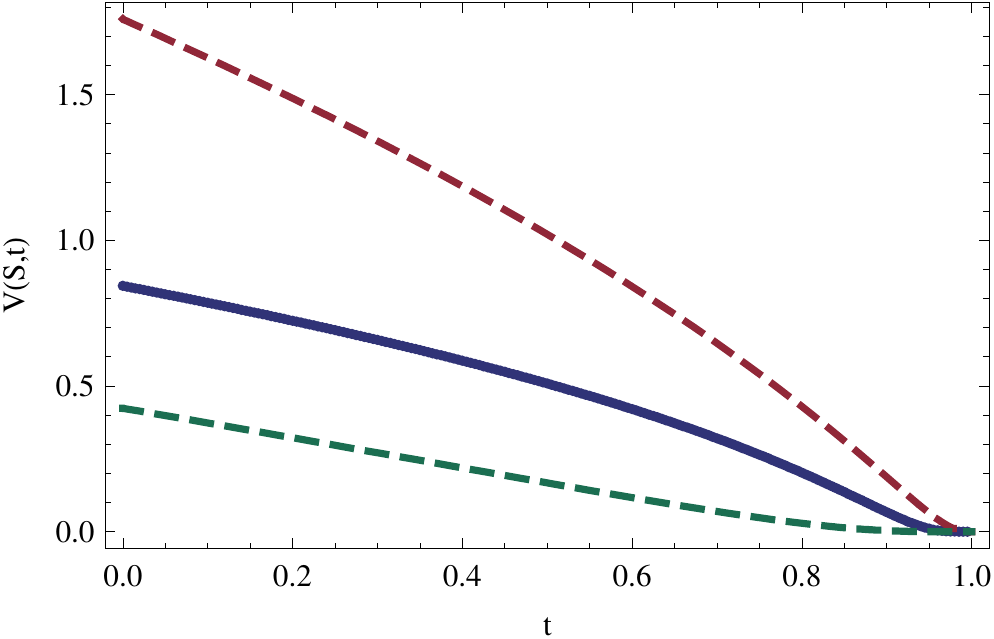}

$S=20$ \hglue 6truecm $S=23$
\medskip

\includegraphics[width=0.48\textwidth]{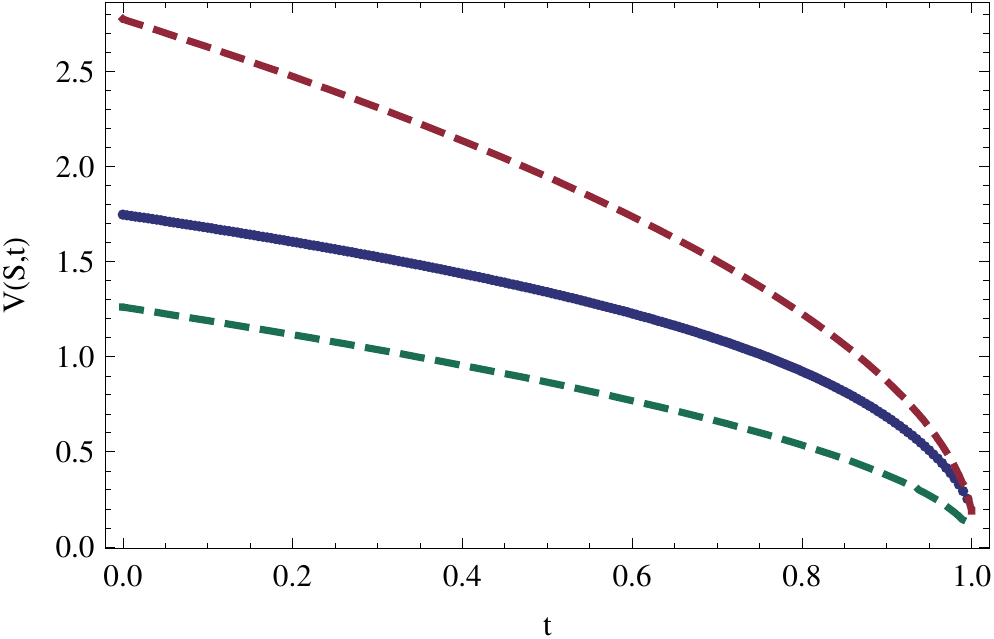}

$S=25\equiv E$

\caption{
The call option price $V(S,t)$ as a function of time $t\in[0,T]$ for $S\in\{20,23, 25\}$.
}
\label{fig:nonlinearTC-SolutioninTime}
\end{center}
\end{figure}

The dependence of the call option price on time $t\in[0,T]$ for $S \in \{20,23,25\}$ with $E=25$ is shown in Figure~\ref{fig:nonlinearTC-SolutioninTime}. We can also see that the price converges to zero  at expiration for $S\le E$.

\subsection{Some practical implications in financial portfolio management}

Numerical results obtained in the previous subsection can be interpreted from the financial portfolio management point of view. For example, temporal behavior of the call option price shown in Fig.~\ref{fig:nonlinearTC-SolutioninTime} for the underlying asset price $S=E$ indicates that the option price given by the solution $V_{vtc}(S,t)$ of the nonlinear variable transaction cost model is closer to the lower bound $V_{\sigma_{min}}(S,t)$ for initial times $t\approx 0$. But in later times when $t\to T$ the price $V_{vtc}(S,t)$  is closer to the upper bound $V_{\sigma_{max}}(S,t)$. It can be interpreted as follows: at the beginning of the contract the portfolio manager need not perform many transaction in order to hedge the portfolio. The transaction costs per one transaction is equal to $C_0$. On the other hand, when the time $t$ approaches expiration $T$ then it is necessary to make frequent rearrangements of the portfolio and so the traded volume of assets increases. Hence the investor pays discounted lower transaction costs value $\underline{C}_0$ per one transaction of the short positioned underlying asset. Consequently, the option price is higher. 

The comparison principle $V_{\sigma_{min}}(S,t) \leq V_{vtc} (S,t) \leq V_{\sigma_{max}}(S,t)$ (see also Table~\ref{tab:bid_price}) has the following practical implication: if the transaction cost $C(\xi)$ per unit share depends on the volume of traded shares $\xi$ and it belongs to the interval $[\underbar{C}_0 \le C \le C_0]$ the option price $V_{vtc}$ can be estimated from above (below) by the option price corresponding to the constant transaction costs  $\underbar{C}_0$ ($C_0$).

\section*{Conclusions}

In this paper we have analyzed a nonlinear generalization of the Black--Scholes equations arising when options are priced under variable transaction costs for buying and selling underlying assets. The mathematical model is represented by the fully nonlinear parabolic equation with the diffusion coefficient depending on the second derivative of the option price. We have investigated properties of various realistic variable transaction costs functions. Furthermore, for a general class of nonlinear Black--Scholes equation we have developed a transformation technique, by means of which the fully nonlinear equation can be transformed into a quasilinear parabolic equation. We have proved existence and uniqueness of classical solutions to the transformed equation. Finally, we have presented a numerical approximation scheme and we computed option prices for pricing model under variable transaction costs.

\end{document}